\documentclass[twoside]{article}
\usepackage{fleqn,espcrc2}
\input epsf
\input{epsf.sty}

\title{Supercooling of the high field vortex phase in single crystalline BSCCO}

\author{C. J. van der Beek$^{a}$, S. Colson$^{a}$, M. Konczykowski$^{a}$,
M.V. Indenbom$^{a,b}$}

\address{Laboratoire des Solides Irradi\'{e}s, Ecole Polytechnique, 
91128 Palaiseau, France\\ 
\noindent $^{{\rm b}}$Institute of Solid State Physics, 142432
Chernogolovka, Moscow District, Russia\\}

\begin{document}

\begin{abstract}
Time resolved magneto-optical images show hysteresis 
associated with the transition at the so-called ``second 
magnetization peak'' at $B_{sp}$ in single-crystalline 
Bi$_{2}$Sr$_{2}$CaCu$_{2}$O$_{8+\delta}$. By rapid quenching of the 
high--field phase, it can be made to persist metastably in the sample 
down to fields that are nearly half $B_{sp}$.
\vspace{1pc}
\end{abstract}

\maketitle


Field--tuned pinning--induced order--disorder transitions 
\cite{Cubitt94} in the vortex 
lattice in type II superconductors have received much attention 
\cite{Wordenweber86II,Bhattacharya93,Khaykovich97II,Kokkaliaris99},  
for the insight they bring to the statistical mechanics of 
elastic manifolds in a random potential, but also because the 
plastically deformed disordered state is that which is most frequently 
encountered in practice, and which determines the high critical 
currents in technological superconductors. 
The transition to the disordered vortex state is particularly 
pronounced in layered materials such as 
Bi$_{2}$Sr$_{2}$CaCu$_{2}$O$_{8+\delta}$ (BSCCO), in which the very 
small vortex line tension allows an optimal adaptation of the disordered 
vortex lattice to the local ``pinscape''; hence, the transition is 
accompanied by a sharp and spectacular increase of the screening 
current $j$ related to pin-

\begin{figure}[h]
    \vspace{-14pt}
    \centerline{\epsfxsize 6.8cm \epsfbox{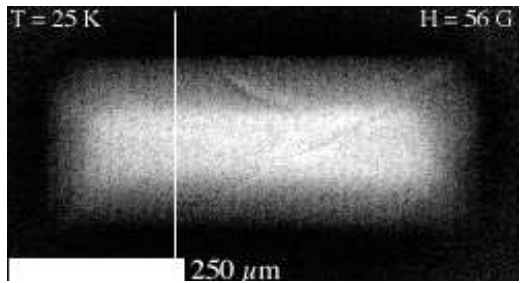}}
    \vspace{-18pt}
\caption{Magneto-optical image of the flux density distribution on 
the  surface of the BSCCO crystal at $T = 25$ K, after the 
magnetic field was rapidly decreased (rise time $< 100$ $\mu$s) from 600 
G to 56 G. The image was taken 0.24 s after the target field was 
reached. }
\label{fig:Fig1}
\end{figure}

\noindent ning, leading to the so-called ``second 
magnetization peak'' (SMP) \cite{Chikumoto92II,Khaykovich97II}. 
This very marked feature has lead to speculations that the SMP represents a transition 
that is peculiar to layered superconductors. Here, we show that 
the transition at the SMP in BSCCO is accompanied by hysteresis and 
the persistence of the metastable high-field (disordered) vortex 
state at inductions $B$ much smaller than $B_{sp}$, at which the 
transition occurs when the system is near thermodynamic equilibrium. 


The presence of the low--field (ordered) and 
high--field (disordered) vortex states in the sample is 
detected by direct imaging of the flux density distribution on the 
sample surface using the magneto-optical technique. The different 
vortex states can then be identified from the different critical 
current density which is related to the local induction gradient 
$\partial B / \partial x$. For the experiments, we 
choose an optimally doped 
BSCCO single crystal of size $630 \times 250 \times 35$ $\mu$m$^{3}$, 
grown by the travelling solvent floating zone technique. The sample was 
covered by a magnetic garnet indicator film with in-plane anisotropy and cooled 
using a continuous flow cryostat. Magnetic fields $H_{a}$ of up to 
600 G could be applied using a split-coil electromagnet surrounding 
the cryostat. The magnet power supply and simultaneous data 
acquisition were controlled using a two-channel synthesized 
wave-generator.


Several types of experiment were performed: (i) rapid field sweeps 
(with different periods $< 10$ s) with synchronized acquisition of 
magneto-optical images at fixed phase (ii) relaxation of the flux 
distribution after a rapid drop of $H_{a}$ from a value 
above or close to $B_{sp}$. An example of the latter is depicted in 
Fig.~1, which shows the flux distribution 0.24 s after the target 
field of 56 G was reached following a quench from $H_{a} = 600$ G (the 
larger intensity corresponds to the greater flux density). Figure 1
shows a bright area of nearly constant $B$ in the crystal 
center, separated from the peripheral low $B$, low $\partial B / 
\partial x$ area by a belt of high flux density gradient. Analysis of the image shows that this high gradient 
is equal in magnitude  to that measured during a slow field ramp for 
$B > B_{sp}$, even though here the local induction during the 
relaxation is (up to 160 G) smaller than $B_{sp}$. The same is 
found when the field is continuously swept at a sufficiently large 
rate. Fig.~2 shows profiles measured during the first and 
second quarter cycles after the application of a triangular waveform 
AC field of amplitude  600 G and frequency  0.1 Hz. The profiles

\begin{figure}[hb]
\vspace{-22pt}
\centerline{\epsfxsize 6.7cm \epsfbox{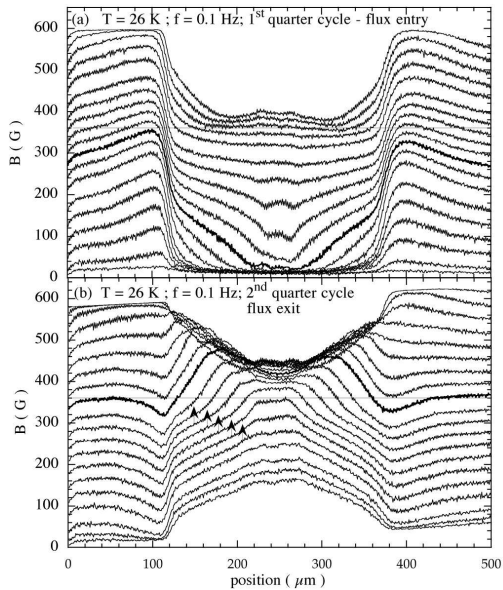}}
\vspace{-18pt}

\caption{Flux profiles along the vertical white line of Fig.~1, taken at 
successive field values during a rapid ramp at
$T = 26$ K. (a) increasing field; 
(b) decreasing field. Arrows indicate the break in slope $\partial 
B / \partial x$ and the phase boundary between the (central) 
disordered vortex state and the (peripheral) ordered vortex lattice. 
Dotted line in either panel indicate $B_{sp}= 360$ G. The irregularities near 
the crystal center are due to the magnetic domain walls 
in the indicator film.}
\label{fig:Fig2}
\vspace{-20pt}
\end{figure}

\noindent are 
characterized by a jump in $B$ at the crystal edge due to the presence 
of the edge barrier current
\cite{Chikumoto92II}, followed by the gradual decrease of 
$B$ in the interior, resulting from the bulk pinning current. 
From the break in the profiles and the appearance of a second 
Bean--like flux front above $B_{sp}$ in panel (a), we determine $B_{sp} = 360$ G 
for $T = 26$ K. Panel (b) shows that 
when $H_{a}$ is rapidly decreased, the relatively large gradient 
present at $B > B_{sp}$ is maintained around a gradually shrinking 
region in the crystal center, while, again, the local induction is 
smaller than $B_{sp}$.

We interpret these observations as being the result of the persistence 
of the metastable high--field vortex state at fields below the 
equilibrium phase boundary. While the appearance of the higher 
$\partial B /\partial x$ at a constant induction $B = B_{sp}$ during 
upward field ramps (Fig.~2(a)), independent of sweep rate, indicates 
that the order-disorder transition at the SMP is a thermodynamic phase 
transition, the observation of a persistent metastable state suggests 
that it is first order. The consequences  are 
multiple. First, the phenomenology of the SMP in BSCCO is
entirely equivalent to that in moderately anisotropic 
\cite{Bhattacharya93,Kokkaliaris99} or even isotropic superconductors 
\cite{Wordenweber86II}, {\em i.e.}, one is dealing with  a similar 
 transition in each case. Second, a first order transition at 
the SMP implies that there is no critical point in the phase diagram, 
as previously suggested \cite{Zeldov95II}. Finally, the presence of 
the quenched disordered vortex state at $B < B_{sp}$ will critically 
affect the outcome of dynamical creep and transport measurements at 
fields near the transition.

\end{document}